\documentstyle[12pt]{article}
\def \half { {1 \over 2}}
\def \d {\delta}

\def \ds {\not \! \partial}
\def \ks {\not \! k}
\def\overlay#1#2{\setbox0=\hbox{#1}\setbox1=\hbox to \wd0{\hss
#2\hss}#1\hskip -2\wd0\copy1}
\def\lsim{\mathrel{\rlap{\lower4pt\hbox{\hskip1pt$\sim$}}
    \raise1pt\hbox{$<$}}}         
\def\gsim{\mathrel{\rlap{\lower4pt\hbox{\hskip1pt$\sim$}}
    \raise1pt\hbox{$>$}}}         
\def\beq{\begin{equation}}
\def\eeq{\end{equation}}
\def\bea{\begin{eqnarray}}
\def\eea{\end{eqnarray}}
\def\nn{\nonumber}

\newcommand\bcdot{{\bf \cdot}}

\newcommand{\bdelta}{\mbox{\boldmath$\delta$}}

\newcommand{\bgamma}{\mbox{\boldmath{$\gamma$}}}
\begin{document}
\begin{center}
{\large \bf Optimized $\bdelta$ expansion for the Walecka model}
\end{center}
\vskip 2.0mm
\begin{center}
G. Krein$^a$, D.P. Menezes$^b$, and M.B. Pinto$^c$
\end{center}
\vskip 2.0mm
\begin{center}
{\it $^a$ Instituto de F\'{\i}sica Te\'orica, Universidade Estadual
Paulista, \\
Rua Pamplona 145,  01405-900 S\~ao Paulo - S.P., Brazil \\
\vskip 0.5cm
$^b$ Departamento de F\'{\i}sica, Universidade Federal de Santa Catarina,\\
88.040-900 Florian\'opolis - S.C., Brazil \\
\vskip 0.5cm
$^c$ Laboratoire de Physique Math\'ematique - Universit\'e de
Montpellier II \\
CNRS-URA 768 - 34095 Montpellier Cedex 05, France }
\end{center}

\vspace{2.5cm}

\begin{abstract}
The optimized $\delta$-expansion is used to study vacuum polarization effects
in the Walecka model. The optimized $\delta$-expansion is a nonperturbative
approach for field theoretic models which combines the techniques of
perturbation theory and the variational principle. Vacuum effects on
self-energies and the energy density of nuclear matter are studied up to
${\cal O}(\delta^2)$. When exchange diagrams are neglected, the traditional
Relativistic Hartree Approximation (RHA) results are exactly reproduced and,
using the same set of parameters that saturate nuclear matter in the RHA, a
new stable, tightly bound state at high density is found.
\end{abstract}
%
\newpage

Although hadrons are not elementary particles, the use of renormalizable
relativistic quantum field models employing pointlike hadrons, such as
the Walecka model~\cite{Wal}, for studying the properties of hadronic matter
at high density and/or temperature is important in several aspects. Perhaps
one of the most important aspects is the necessity of understanding the vacuum
of such models, since one expects that the description of highly excited
matter in terms of such models must breakdown at some scale and then
quark-gluon degrees of freedom must be invoked for a proper treatment of the
system. Quantum fluctuations in the Walecka model have been studied at the
Hartree level~\cite{SerWal}, but severe difficulties arise when
nonperturbative exchange diagrams (Fock diagrams) are considered. Although
Bielajew and Serot~\cite{BielSer} have set up an appropriate framework for the
renormalization of the Hartree-Fock (HF) equations, no explicit calculation
employing this has been performed so far. Bearing in mind the difficulties
associated with the renormalization within the HF approximation we give in
this Letter the first step towards the implementation of a different
nonperturbative approach for studying vacuum fluctuations in hadronic models.
Namely, we use the $\delta$ expansion~\cite{original}; more specifically
we use the {\em optimized linear} $\delta$-{\em expansion}~\cite{linear}.

The standard application of the linear
$\delta$-expansion~\cite{{linear},{njl},{KMNP}} to a theory with
action $S$ starts with an interpolation defined by
\begin{equation}
S(\delta) = (1-\delta)S_0(\mu) + \delta S = S_0(\mu) + \delta [S-S_0(\mu)],
\label{int}
\end{equation}
where $S_0(\mu)$ is the action of a solvable theory. The action
$S(\delta)$ interpolates between the solvable $S_0(\mu)$ (when $\delta=0$)
and the original $S$ (when $\d=1$). Since  $S_0$ is quadratic in the
fields, arbitrary parameters ($\mu$) with mass dimensions are required for
dimensional balance. The evaluation of a physical quantity $P$ is
performed by considering the term $\delta [S-S_0(\mu)]$ as a perturbation
whose order is labeled by $\delta$  which is set to
unity at the end. In practice, the perturbative expansion in powers of
$\delta$ will be truncated at a given order implying that the quantity $P$
will have a residual dependence on the unknown parameters. Since
$\mu$ does not belong to the original theory it will have to be fixed
according to some criterion and different methods have been
proposed. Among them the Principle of Minimal Sensitivity (PMS)~\cite{pms}
offers a particularly attractive way of optimizing the theory by requiring
$P(\mu)$ to be evaluated at the point where it is less sensitive to small
variations of $\mu$: ${\partial P(\mu) / {\partial \mu}}= 0$.
With this variational procedure $\mu$ becomes a function of the original
parameters of the theory yielding non-perturbative results.

The different forms of the $\delta$ expansion have been successfully applied
to many different problems in quantum mechanics, particle
theory, statistical physics and lattice
field theory, and its convergence has been recently
proved for quantum mechanical problems~\cite{liter}. More recently, the
optimized linear $\delta$-expansion was used for applications in $\Phi^4_4$,
QED and Yang-Mills theories~\cite{gromes}. In a recent paper~\cite{KMNP} it
was demonstrated that in truncating the $\delta$ expansion at
${\cal O}(\delta^2)$, and neglecting vacuum effects, one can readily reproduce
the results of the standard self-consistent Dirac-Hartree-Fock
approximation~\cite{SerWal} (in this approximation vacuum effects are also
neglected) for the equation of state of nuclear matter. In this Letter we go
one step further by including vacuum effects to ${\cal O}(\delta^2)$ and
demonstrate that one can reproduce the standard relativistic Hartree results
(RHA). In addition, from the PMS applied to the renormalized energy density of
nuclear matter, a new stable state is found at high density. One of the
stengths
of the $\delta$-expansion is that one can proceed to beyond leading order with
considerably less computational effort than in the traditional HF approximation
since in one deals in this approach with a finite number of Feynman graphs as
implied by the perturbative nature of the calculation.

We start with the Lagrangian density of the Walecka model~\cite{Wal,SerWal}:
\bea
{\cal L}  &=& \nonumber  \bar \psi(i \ds - M+ g_s \phi - g_v \gamma^{\mu}
V_{\mu})\psi + \half (\partial_{\mu}\phi\partial^{\mu}\phi - m_s^2
\phi^2)\nonumber\\
& & - {1 \over 4} F_{\mu \nu}F^{\mu \nu} + \half m_v^2 V_{\mu}V^{\mu}
+ U(\phi,V) + {\cal L}_{CT},
\label {ow}
\eea
where $\psi$ represents the nucleon field operator, $\phi$ and $V_{\mu}$ are
respectively the field operators of the scalar and vector meson and
$F_{\mu \nu}= \partial_{\mu} V_{\nu} - \partial_{\nu} V_{\mu}$. The massive
vector field is coupled to a conserved baryon current, rendering the model
renormalizable in 3+1 dimensions.  In
order to minimize many body effects the term $U(\phi,V)$, which describes
mesonic self interactions, was set to zero in the original work of  Walecka
and we shall keep the same convention here. The Lagrangian
density ${\cal L}_{CT}$ contains all the counterterms needed to render the
model finite which, for the purposes of the present paper, is given by:
\begin{equation}
{\cal L}_{CT} = \sum_{n=1}^{4}\frac{\beta_n}{n!}\phi^n + \zeta_{\rm f} \,
\bar \psi (i\ds - M)\psi + M_c \bar \psi  \psi.
\label{Lct}
\end{equation}
We are primarily interested in studying effects of vacuum fluctuations in
nuclear matter. In particular, we will consider the energy per nucleon,
which is related to the energy density by:
\beq
{\cal E} = \frac{1}{V}\int d^3 x \, \left( <\!\Psi|T^{00}|\Psi\!>
 - <\!{\rm vac}|T^{00}|{\rm vac}\!>\right)={\cal E}^B+{\cal E}^s+{\cal E}^v,
\label{Edens}
\eeq
where $|\Psi\!>$ is the interacting ground-state of nuclear matter,
$|{\rm vac}\!>$ is the vacuum state (zero density), $T^{00}$ includes the
piece corresponding to ${\cal L}_{CT}$. The terms ${\cal E}^B$,
${\cal E}^s$ and ${\cal E}^v$ refer respectively to baryon, scalar-meson,
and vector-meson contributions, which are given by:
\begin{eqnarray}
{\cal E}^B &=& -i\int\frac{d^4k}{(2\pi)^4}\left\{{\rm Tr}
\left[\gamma^0 k^0 - ({\not\!k}-M)\right]S(k)\right\} - {\cal E}^B_{VEV} +
{\cal E}^B_{CT},
\nonumber\\
\label{EB}
{\cal E}^s &=&\frac{1}{2}\frac{g_s^2}{m_s^2}
\left[\int\frac{d^4k}{(2\pi)^4}{\rm Tr}S(k)\right]^2 -
g_s^2\int\frac{d^4k}{(2\pi)^4} \frac{d^4q}{(2\pi)^4}
{\rm Tr}\left[S(k+q)S(q)\right] \Delta_s(k^2) \nn\\
&& \times \,  \left\{ \left[\frac{1}{2}(k^2-m_s^2)\Delta_s(k^2)-1\right]
-{\left(k^0\right)}^2 \Delta_s(k^2)\right\} - {\cal E}^s_{VEV} +
{\cal E}^s_{CT},\nonumber\\
\label{Es}
{\cal E}^v &=& - \frac{1}{2}\frac{g_v^2}{m_v^2}\left[\int\frac{d^4k}{(2\pi)^4}
{\rm Tr}\gamma^0 S(k)\right]^2 + g_v^2\int\frac{d^4k}{(2\pi)^4}
\frac{d^4q}{(2\pi)^4}{\rm Tr}\left[\gamma_\lambda S(k+q)\gamma^\lambda S(q)
\right]\Delta_v(k^2)\nonumber\\
&& \times \, \left\{\left[\frac{1}{2}(k^2-m_v^2)\Delta_v(k^2)-1\right]
- {\left(k^0\right)}^2 \Delta_v(k^2)\right\} - {\cal E}^v_{VEV} +
{\cal E}^v_{CT},
\label{Ev}
\end{eqnarray}
where $\Delta_m(k), m=s,v$ are the Fourier transforms of Green's functions of
the Klein-Gordon operator. It is important to note that we are not using the
nucleon equation of motion and, therefore, the above expressions differ from
the usual ones~\cite{SerWal}. The reasons for not using the nucleon equation
of motion will be discussed below. The terms ${\cal E}^{B,s,v}_{CT}$
will be considered in renormalization of the energy density. In obtaining
these expressions, we have eliminated the
meson field operators $\phi$ and $V^{\mu}$ in favor of the nucleon field
operators by integrating the meson Euler-Lagrange equations.

The problem of calculating the energy density consists therefore in finding
the interacting nucleon propagator in medium. The first step towards the
calculation of the propagator with the $\delta$ expansion is to define the
interpolated Lagrangian. According to Eq.~(\ref{int}), one has:
\begin{equation}
{\cal L}(\delta) = (1 - \delta) \, {\cal L}_0 + \delta \, {\cal L} \\
= {\cal L}_0 + \, \delta \, \bar \psi(g_s \phi -
g_v \gamma^{\mu}V_{\mu} + \mu)\psi + {\cal L}_{CT}(\delta).
\label{inter}
\end{equation}
where we have chosen ${\cal L}_0$ to be:
\beq
{\cal L}_0=\bar \psi\left(i\gamma_{\mu}\partial^{\mu}-\Omega\right)\psi
+\frac{1}{2}(\partial_{\mu}\phi\partial^{\mu}\phi-m_\sigma^2\phi^2)-
\frac{1}{4}F_{\mu\nu}F^{\mu\nu}+\frac{1}{2}m_\omega^2V_{\mu}V^{\mu}\;,
\label{L0Wal}
\eeq
where $\Omega \equiv M+\mu$. Notice that the $\delta$-expansion interpolation
could also have been done in the mesonic sector. However, we have chosen
to eliminate the mesonic fields by integrating their equations of motion; in
this way mesonic self-energies are automatically taken into account. Of course,
the integration in favor of the nucleon fields can be done only in the absence
of mesonic self-interactions [Note that since we have set $U(\phi,V)=0$
the mesonic self-interactions appear in
${\cal L}_{CT}$ only]. This leaves us with only one unknown parameter, $\mu$,
which will be fixed by the PMS condition applied to the energy density. In
short, one is performing a variational calculation of the energy density with
an ``educated guess" for the nucleon propagator which can be improved
perturbatively as higher powers of $\delta$ are considered. Since one is
looking for the propagator that leaves the energy stationary, one is not
allowed to substitute the nucleon equation of motion in the expression of
the energy.  We could also interpolate kinetic energy terms, as
in Ref.~\cite{gromes}, but we prefer to start with the simplest interpolation.
Such generalizations will be examined in forthcoming publications.

The fact that the divergent part of a physical quantity $P(\mu)$
calculated with this interpolated Lagrangian will be $\delta$
and $\mu$ dependent implies that the coefficients appearing in the counterterm
Lagrangian ${\cal L}_{CT}(\delta)$ will also depend on both parameters. The
interpolated ${\cal L}_{CT}(\delta)$ has the same field-operator structure
as the original ${\cal L}_{CT}$ of Eq.~(\ref{Lct}), with $\delta$- and
$\mu$-dependent coefficients. The explicit $\delta$ and $\mu$ dependence is
not important since this dependence will appear automatically in the process of
fixing the value of the coefficients in the renormalization process.

Next the strategy consists in calculating the interacting propagator, which
is obtained by inverting Dysons's equation $S^{-1}(p)=S^{0^{-1}}(p)-\Sigma(p)$,
where the self-energy $\Sigma(p)$ is calculated as a perturbation expansion in
powers of $\delta$, using for the ``non-interacting" propagator, $S^0$, the one
corresponding to ${\cal L}_0$. $S^0$ can be split in the usual way as
$S^0(k)=S^0_F(k) + S^0_D(k)$, where $S^0_F(k)$ and $S^0_D(k)$ are the Feynman
and density dependent parts for quasi-particles of mass $\Omega$ and energy
$E_{\Omega}=( {\vec k}^2 + \Omega^2)^{1/2}$. Inversion of Dyson's equation is
standard~\cite{SerWal}. First, one notes that the in-medium $\Sigma(p)$ is of
the general form~\cite{SerWal}: $\Sigma(k) = \Sigma^s(k)-\gamma^0\Sigma^0(k)
+{\bgamma}{\bcdot} {\bf k} \Sigma^v(k)$. Then, one defines the auxiliary
quantities:
\begin{eqnarray}
&&\Omega^*(k) = \Omega +\Sigma^s(k)\hspace{2.5cm}{\bf k}^* = {\bf k}\left[1+
\Sigma^{v}(k)\right],\nonumber\\
&&E^*(k) = \left[{\bf k}^{*2}+\Omega^{*2}(k)\right]^{1/2},\hspace{1.0cm}
k^{* \mu} = k^\mu + \Sigma^\mu (k)=\left[k^0+\Sigma^0(k),
{\bf k}^*\right].
\label{Aux}
\end{eqnarray}
The propagator can be written as $S(k) = S_F(k) + S_D(k)$, with:
\begin{equation}
S_F(k) = \frac{1}{\not\! k^* - \Omega^*(k)+i\epsilon},\hspace{1.0cm}
S_D(k) = i\pi \frac{\not\! k^* + \Omega^*(k)}{E^*(k)}
\delta\left(k^0-E(k)\right)\theta\left(k_F-|{\bf k}|\right)\;,
\label{SFD}
\end{equation}
where $E(k)$ is the single-particle energy, which satisfies
$E(k)=\left[E^*(k)-\Sigma^0(k)\right]_{k^0=E(k)}$.
Note that we have assumed that the nucleon propagator has simple
poles with unit residue. Within the approximation scheme we will work in
this paper, this assumption is satisfied, as can be seen below.

The renormalization procedure in the $\delta$ expansion follows closely the
usual perturbative renormalization program, with small, however important,
differences. Hence, in the following we shall repeat some standard
textbook material which we feel essential for the appreciation of the
differences.  We start with the problem in vacuum ($k_F =0$), in which case
$S^0(k)=S^0_F(k)$. We perform our calculations in $2 \omega=4-2\epsilon$
dimensions using dimensional regularization techniques~\cite{co}.
To ${\cal O}(\delta)$,  the self-energy is simply $\Sigma^{(1)}_F(k)= - \delta
\, \mu$, which arises from the bilinear term $\d \,\mu\,\bar\psi\psi$. To
O($\delta^2$), one has in principle tadpole and exchange contributions:
\begin{equation}
\Sigma^{(2)\rm tad}_F(k) = i \delta^2 \left(\frac{g_s}{m_s}\right)^2
\int \frac{d^4q}{(2\pi)^4} \, {\rm Tr}\left[S^0_F(q)\right]
- i \delta^2 \left(\frac{g_v}{m_v}\right)^2\int \frac{d^4q}{(2\pi)^4} \,
\gamma_\mu{\rm Tr}\left[\gamma^\mu S^0_F(q)\right],
\label{Tad2}
\end{equation}
and
\begin{equation}
\Sigma^{(2)\rm exch}_F(k)= i \delta^2 g_s^2 \int\frac{d^4 q}{(2\pi)^4} \,
S^{(0)}_F(q)\Delta_s(k-q)-i \delta^2 g_v^2 \int\frac{d^4 q}{(2\pi)^4} \,
\gamma_\mu S^{(0)}_F(q)\Delta_v(k-q)\gamma^\mu.
\label{Exch2}
\end{equation}

The vector meson tadpole contribution vanishes after taking the trace and
performing the integral, whereas scalar-meson tadpole contains a finite
and a divergent part. The renormalization of this is performed
using the term $\beta_1(\delta)\phi $ from ${\cal L}_{CT}(\delta)$ in
Eq.~(\ref{Lct}). Its contribution to the self-energy is
$\Sigma_{\beta_1}=\beta_1(\delta) \, g_s \, \Delta_s(0)$. In order to produce
a stable vacuum, $\beta_1(\delta)$ is chosen so as to eliminate the divergent
and finite parts. Any finite piece must be canceled, since such a term would
contribute to the energy of the vacuum, which in turn could be lowered without
bound.

The exchange parts of the self-energy are given by:
\beq
\Sigma^{(2)\rm exch}_F(k) = - \delta^2 \left[\ks a(k^2) - b(k^2)\right],
\label{Sigexch}
\eeq
where
\begin{eqnarray}
a(k^2) &=& \frac{g_s^2}{(4\pi)^2} \left[\frac{1}{\epsilon}
- I_0^s(k^2) \right] +  \frac{2g_v^2}{(4\pi)^2} \left[\frac{1}{\epsilon}
- I_0^v(k^2) \right] + C_a,\nonumber\\
\label{a}
b(k^2) &=& -\frac{g_s^2 \Omega}{(4\pi)^2}\left[
\frac{1}{\epsilon} - I_1^s(k^2) \right]
+ \frac{4 g_v^2 \Omega}{(4\pi)^2}\left[\frac{1}{\epsilon} -
I_1^v(k^2)\right ] + C_b,
\label {b}
\end{eqnarray}
where $C_a$ and $C_b$ are irrelevant constants and $I^m_i(k^2),\;\;i=1,2$ are
the integrals:
\begin{equation}
I_i^m(k^2) = \int_0^1 (d \alpha)_i \ln \left [ \frac{k^2(\alpha-1)\alpha
+ (1-\alpha)\Omega^2 + \alpha m^2_m)}{ m_m^2} \right ] \;\;\;\;,
\end{equation}
where $(d \alpha)_0 = \alpha d\alpha$ and $(d \alpha)_1 = d\alpha$ , with
$\alpha$ being the Feynman parameter.

The renormalization of the exchange self-energy is carried out with the last
two terms of Eq,~(\ref{Lct}), which contribute to the self-energy as
$\Sigma^{\rm CT}_F(k)= - M_c(\delta) + \zeta_{\rm f}(\delta)(\ks - M)$, and
the renormalized vacuum contribution to the self-energy,
$\Sigma^{R}_{F}(k)$, up to ${\cal O}(\delta^2)$ is given by
$\Sigma^{R}_{F}(k) = - \delta\,\mu + \Sigma^{(2)\rm exch}_F(k) +
\Sigma^{\rm CT}_F(k)$.

Generally, counterterms are composed of a divergent part which completely
eliminates the poles and of an arbitrary finite part which is fixed according
to a chosen renormalization scheme~\cite{co}. Here we shall consider the
parameters $g_m$,$m_m$ and $M$ to be ``the" renormalized physical couplings
and masses. This choice amounts to the on-mass shell renormalization scheme
in which the counterterms remove {\it both} divergent and finite contributions
from loop corrections to measurable amplitudes. Within this renormalization
scheme the finite parts of both counterterms are fixed by the renormalization
conditions:
\begin{equation}
S_F^{-1}(k)|_{{\not\! k} = M} = \left[ S^{0^{-1}}_F(k)-\Sigma^{R}_{F}(k)
\right]\Big |_{{\not\! k} = M}=0\hspace{1,0cm}{\rm and}\hspace{1.0cm}
\frac{\partial S_F^{-1}(k)}{ \partial {\not\! k} }
\Bigg |_{{\not\! k} = M}=1.
\label{RenCond}
\end{equation}
Application of these conditions leads to the following expression
for the renormalized self-energy in vacuum:
\begin{equation}
\Sigma^{R}_{F}(k) = A(k^2) {\not\! k} - B(k^2),
\label{RenorSig}
\end{equation}
where the functions $A(k^2)$ and $B(k^2)$ are free of divergencies and given
by:
\begin{equation}
A(k^2)=-\delta^2 \, \left[ \bar a(k^2) - 2M\bar c(M^2) \right],\hspace{1.0cm}
B(k^2)= \mu - \delta^2 \, \left[ \bar b(k^2) - 2M^2 \bar c(M^2)\right],
\label{AandB}
\end{equation}
with $\bar a(k^2) = a(k^2) - a(M^2)$, $\bar b(k^2) = b(k^2) - b(M^2)$, and
$\bar c(M^2) = M \bar a'(M^2)-\bar b'(M^2)$, where the prime
denotes a derivative with respect to the argument. It is easy to check that
Eqs.~(\ref{RenCond}) are indeed satisfied by Eqs.~(\ref{RenorSig}-\ref{AandB}).

In principle one can proceed to higher orders in $\delta$ and include, for
example, the important vertex corrections. The renormalization at higher
orders introduces no extra complications as compared to the usual perturbative
renormalization.
Instead of going to higher orders, we consider next the problem of the energy
density, where new divergencies arise and extra renormalization is required.
In order to simplify the discussion, we shall consider here only direct terms
and then compare the results with the ones obtained within the RHA.

When considering direct terms only, up to ${\cal O}(\delta^2)$ one obtains the
following relations for the auxiliary quantities defined in Eq.~(\ref{Aux}):
\begin{eqnarray}
&&\Omega^*(k)=\Omega - \delta \mu - \delta^2 4 \frac{g_\sigma^2}
{m_\sigma^2} \int_0^{k_F}\frac{d^3q}{(2\pi)^3}\frac{\Omega}{E_\Omega(q)},\\
&& k^{0\,*} =  k^{0} - \delta^2 4 \frac{g_\omega^2}
{m_\omega^2}\int_0^{k_F}\frac{d^3q}{(2\pi)^3},\hspace{1.0cm}
{\bf k}^{*} = {\bf k}
\label{Dir*}
\end{eqnarray}
{}From these, one constructs the nucleon propagator which is
then used in the expression for the energy density.
We start with the baryon contribution to the energy density:
\begin{eqnarray}
{\cal E}^B &=& - { 3{\Omega^{*}}^4 \over {8 \pi^2}} \left [
{1 \over \epsilon } + \psi (3) +
\ln \left ( {4 \pi\eta^2 \over {{\Omega^{*}}^2}} \right ) \right ]
+{ M{\Omega^{*}}^3 \over {2 \pi^2}} \left [ {1 \over \epsilon } +
\psi (2) + \ln \left ( {4 \pi \eta^2 \over {{\Omega^*}^2}} \right ) \right ]
\nonumber\\
&+& 4 \int_0^{k_F} {d^3 {\vec k} \over (2\pi)^3}
{ { {\vec k}^2 + M \Omega^{*}} \over E^{*}(k)} - {\cal E}^B_{VEV} +
{\cal E}^B_{CT}.
\end{eqnarray}
Note that the divergent and finite parts are different from the usual
ones~\cite{SerWal} because we have not used the nucleon field equation of
motion. Carrying out the VEV subtraction, the remaining
divergencies are eliminated by the counterterm $\sum_{n=2}^4 (\beta_n/n!)
\phi^n$ in Eq.~(\ref{Lct}). Thus, the finite contribution to the baryonic part
of the energy density is given by (after taking $\delta=1$):
\begin{equation}
{\cal E}^B =  \int_0^{k_F} {d^3 {\vec k} \over (2\pi)^3}
{ { {\bf k}^2 + M \Omega^* } \over E^*(k) } + \Delta_B,
\label{EBfin}
\end{equation}
where
\begin{eqnarray}
\Delta_B &=& {1\over {4\pi^2}}\left\{
\ln \left ( {\Omega^* \over M} \right)
\left [ 3(\Omega^{*})^4 - 4M (\Omega^{*})^3 \right ]\right.\nonumber\\
&+& \left.   M^3(\Omega^*-M)
- { 1 \over 2} M^2(\Omega^*-M)^2 - {17 \over 3}M(\Omega^*-M)^3 -
{21 \over 4} (\Omega^*-M)^4 \right\}.
\label{DeltaB}
\end{eqnarray}

Proceeding in the same way for the scalar meson contribution, we obtain:
\begin{equation}
{\cal E}^s = -\frac{m_s^2}{2g_s^2} \left(4\frac{g_s^2}{m_s^2}
\int \frac{d^3k}{(2\pi)^3} \frac{\Omega^*}{E^*(k)}
- \Delta_{S}\right)^2
\label{Esca}
\end{equation}
where
\begin{equation}
\Delta_{S} = \frac{g_s^2}{m_s^2}\frac{1}{\pi^2}
\left[{\Omega^*}^3 \ln \left(\Omega^*\over M\right) - M^2(\Omega^*-M) -
{5\over 2} M(\Omega^*-M)^2 - {11\over 6} (\Omega^*-M)^3 \right].
\label{es}
\end{equation}

Finally, there are no divergencies with the vector meson contribution, it is
given by:
\begin{equation}
{\cal E}^v = \half {g^2_v \over m_v^2}
\left[ {2 \over 3 \pi^2} k_F^3 \right ]^2 .
\label{Evv}
\end{equation}

The determination of the unknown parameter $\mu$ follows from the PMS
applied to the energy density:
\begin{equation}
\frac{d{\cal E}}{d \mu} = \frac{d{\cal E}}
{d \Omega^*} \frac {d \Omega^*}{d \mu} = 0 \Rightarrow
\frac {d {\cal E} }{d \Omega^*} = 0,
\label{PMSE}
\end{equation}
where we used the fact that ${d \Omega^*}/{d \mu} \neq 0$. Differentiation of
${\cal E}$ with respect to $\Omega^*$ leads to the following self-consistent
equation for the effective nucleon mass $\Omega^*$:
\begin{equation}
\left\{\left(M-\Omega^*\right)-\left[4 \frac{g_s^2}{m_s^2}
\int \frac{d^3k}{(2\pi)^3}
\,\frac{\Omega^*}{E^*(k)}-\Delta_S\right]\right\}\left[4\frac{g_s^2}{m_s^2}
\int \frac{d^3k}{(2\pi)^3}\,
\frac{{\bf k}^2}{E^{*3}(k)}-\frac{d \Delta_{S}}{d\Omega^*}\right]=0.
\label{self}
\end{equation}
To arrive at this result we made use of the following identity:
\begin{equation}
\frac{d \Delta_B}{d\Omega^*}=(\Omega^*-M)\frac{m_s^2}{g_s^2}
\frac{d \Delta_{S}}{d\Omega^*}
\end{equation}
Clearly, Eq.~(\ref{self}) admits two solutions. The one that follows from
the vanishing term in curly braces corresponds to the usual RHA
self-consistent solution. It is not difficult to show that when this is
substituted into the equation for the energy density, one obtains the RHA
energy density:
\begin{equation}
{\cal E}= \frac{g_v^2}{2m_v^2}\left(\frac{2}{3\pi^2}k_F^3\right)^2 +
\frac{m_s^2}{2g_s^2}(\Omega^*-M)^2 + 4 \frac{g_s^2}{m_s^2}
\int \frac{d^3k}{(2\pi)^3} \,{E^*(k)} + \Delta{\cal E}_{VF}
\label{EdenCompl}
\end{equation}
where $\Delta{\cal E}_{VF}$ is the energy density corresponding to vacuum
fluctuations, given by:
\begin{eqnarray}
\Delta{\cal E}_{VF} &=& \frac{1}{4\pi^2}
\left[{-\Omega^*}^4 \ln \left(\Omega^*\over M\right) + M^3(\Omega^*-M)
\right.\nonumber\\
&+& \left. {7\over 2} M^2(\Omega^*-M)^2 + {13\over 3} (\Omega^*-M)^3
+ {25\over 12}(\Omega^*-M)^4\right].
\label{EVF}
\end{eqnarray}
With the parameter set~\cite{SerWal}, $g_s^2=62.89$, $g_v^2=79.78$, $m_s= 550$
MeV, $m_v= 783$ MeV and $M=939$ MeV, nuclear matter is saturated at $k_F =
1.42\; {\rm fm}^{-1}$, with a binding energy of $16$ MeV per nucleon.

The other solution to Eq.~(\ref{self}) is:
\begin{equation}
\frac{d \Delta_{S}}{d\Omega^*} = 4\frac{g_s^2}{m_s^2}
\int \frac{d^3k}{(2\pi)^3}\,\frac{{\bf k}^2}{E^{*3}(k)}.
\label{unusual}
\end{equation}
Note that the solutions of this equation are {\em independent} of the
parameters of the model; the effective nucleon mass $\Omega^*$ depends only
on $M$ and $k_F$. The numerical solution of this equation gives an effective
nucleon mass that is an {\em increasing} function of the nuclear density.
Moreover, when this solution is substituted into the original expression for
the energy density, and using the same parameters as the ones of the RHA,
we find a new saturation point for nuclear matter at $k_F \sim 5.3\;
{\rm fm}^{-1}$, which corresponds to a density of approximately 50 times
the one of normal nuclear matter, with a binding energy of the order of 570
MeV per nucleon. Of course, if the parameters are changed, these numbers will
change. However, we do not pursue the discussion on this further because
exchange terms should be included for consistency (they are of the same order
in $\delta$ as the direct ones), although one should notice that in all known
cases where exchange graphs are included their effect can be absorbed by
a readjustment of parameters~\cite{SerWal}.

When one considers the exchange diagrams in the energy density, the
renormalization becomes more difficult, but not more complicated than the
usual renormalization in the loop expansion~\cite{FPS}. In particular, if
instead of inverting Dyson's equation for obtaining the propagator from the
self-energy one uses the same approach as Chin~\cite{Chin} for treating
the exchange diagrams, the renormalization is indeed simplified.

We have given the first step towards the implementation of the
$\delta$-expansion for studying vacuum effects in effective hadron field
theories. We demonstrated how the renormalization program can be implemented
within the $\delta$ expansion. We also calculated the renormalized energy
density of nuclear matter at ${\cal O}(\delta^2)$ by neglecting exchange
diagrams, and found that the usual RHA approximation is readily reproduced. In
addition, a new state of nuclear matter is found from the PMS applied to the
renormalized energy density. Concluding, we believe that the results are very
encouraging for future investigations concerning the vacuum of hadronic models.

This work was partially supported by CNPq and FAPESP
(contract \# 93/2463-2). The authors would like to thank Marina Nielsen
and MBP would like to thank P. Grang\'e and G. Mennessier for helpful
discussions.

\end{document}